\documentclass[12pt]{article}
\usepackage{amsmath,amssymb}
\usepackage{graphicx}
\usepackage{cite,fancyhdr}

\newcommand{\Slash}[1]{{\ooalign{\hfil#1\hfil\crcr\raise.167ex\hbox{/}}}}

\newcommand{\beq}{\begin{equation}}  \newcommand{\eeq}{\end{equation}}
\newcommand{\bef}{\begin{figure}}  \newcommand{\eef}{\end{figure}}
\newcommand{\bec}{\begin{center}}  \newcommand{\eec}{\end{center}}
  
\newcommand{\laq}[1]{\label{eq:#1}}  

\newcommand{\Eq}[1]{Eq.(\ref{eq:#1})}

\newcommand{\eq}[1]{(\ref{eq:#1})}

\newcommand{\ab}[1]{\left|{#1}\right|}

\newcommand{\SU}[1]{{\rm SU{#1} } }

\def\({\left(}
\def\){\right)}

\def\O{\mathcal{O}}
\def\U{\mathop{\rm U}}

\newcommand{\OR}{~{\rm or}~}
\newcommand{\AND}{~{\rm and}~}

\newcommand{\GEV}{ {\rm ~GeV} }
\newcommand{\TEV}{ {\rm ~TeV} }
\def\o{\over}
\def\a{\alpha}

\def\d{\delta}

\def\f{\phi}
\def\g{\gamma}

\def\k{\kappa}
\def\l{\lambda}
\def\m{\mu}
\def\n{\nu}

\def\s{\sigma}

\def\D{\Delta}

\def\ol{\overline}

\def\*{\dagger}
\usepackage[affil-it]{authblk}
\usepackage[left=2cm,right=2cm,top=2cm,bottom=2cm,a4paper]{geometry}

\begin{document}
\begin{titlepage}
\begin{center}
\setcounter{footnote}{0}
\setcounter{figure}{0}
\setcounter{table}{0}

{\Large\bf 
Muon $g-2$ at multi-TeV muon collider
}

\vskip .75in

{ \large    Wen Yin\,$^{a}$ and    Masahiro Yamaguchi\,$^{b}$},

\vskip 0.25in

\begin{tabular}{ll}
$^{a}$ &\!\! {\em Department of Physics, University of Tokyo, Tokyo 113-0033, Japan} \\
& {\em Department of Physics, KAIST, Daejeon 34141, Korea,} \\[.3em]
$^{b}$ &\!\! {\em Department of Physics, Tohoku University, }\\
& {\em Sendai, Miyagi 980-8578, Japan}\\[.3em]

& {\em }

\end{tabular}

\begin{abstract}
 The long-standing discrepancy of muon $g-2$ is a hint of new physics beyond the standard model of particle physics. 
In this letter we show that heavy new physics contribution can be fully tested at a muon collider with center-of-mass energy up to $\O(10)\TEV.$ 
Even if there is no new particle in this energy range, one can measure the $g-2$ directly via the channel to a Higgs boson and a monochromatic photon.
\noindent
\end{abstract}

\end{center}
\end{titlepage}
\setcounter{footnote}{0}
\setcounter{page}{1}

\section{Introduction}
The long-standing discrepancy of muon anomalous magnetic moment ($g-2$) is the leading candidate suggesting new physics that couples to the standard model (SM). 
The muon $g-2$ anomaly indicates the more than 3 $\sigma$ level deviation of
\beq
\laq{g2m}
\D a_\mu = a_\mu^{\rm EXP} - a_\mu^{\rm SM} = (27.4 \pm 7.3) \times 10^{-10},
\eeq
where $a_\m^{\rm SM}$ is the SM prediction of the muon $g-2$ from the so-called R-ratio approach~\cite{Davier:2017zfy, Keshavarzi:2018mgv} (see  also \cite{Keshavarzi:2019abf, Aoyama:2020ynm}) and 
$a_\mu^{\rm EXP}$ is its experimental result~\cite{Bennett:2006fi, Roberts:2010cj}.  
The ongoing experiment E989 at Fermilab \cite{Grange:2015fou} and the upcoming one at J-PARC \cite{Mibe:2010zz} may significantly increase the accuracy of the experimental value. 
On the other hand,  the leading order vacuum polarization contribution was recently calculated by Borsanyi et al  in the Lattice QCD\,\cite{Borsanyi:2020mff} and was argued to resolve the tension, although the result is still on debate~\cite{Crivellin:2020zul, Keshavarzi:2020bfy}. Therefore, in order to confirm the discrepancy, efforts are being made on both experimental and theoretical sides. 
In this Letter, on the contrary, we propose a high energy test of the $g-2$. This is not  bothered by the QCD non-perturbative effect. 

On the model-building side, if an ultra-violet (UV) theory generates the $g-2$ anomaly, the resulting low energy effective theory should have 
the muon $g-2$ operator given by 
\beq
\laq{g-2}
\Delta {\mathcal L}_{\rm eff}\supset { e \D\a_\m \o 4 m_\m} \ol{\m}  \s_{\m\n}   F^{\m \n}    {\m}  
\eeq
Here $m_\mu\simeq 0.11\GEV$ is the muon mass, where $\s_{\mu \nu}\equiv i[\g_\m ,\g_\n]/2, $ and $F^{\m\n}$ is the field strength of photon. 
 In this UV explanation, we may either have new states much heavier than $m_\m$ to generate \eq{g-2},  
or have strongly-coupled theory at the high energy since \eq{g-2} is a higher dimensional term. 
On the contrary, the $g-2$ may be also explained due to a weakly-coupled light particle, below or around the muon mass, i.e. infra-red (IR) explanation.

In this letter, we point out that the UV explanation scenarios can be fully tested at a muon collider \cite{Ankenbrandt:1999cta, Delahaye:2019omf, Garcia:2020xrp} (see also recent studies for BSM \cite{Costantini:2020stv, Chiesa:2020awd, Han:2020pif, Han:2020uak,Capdevilla:2020qel}) with center-of-mass energy up to $\O(10)\TEV$.
This is because the heavy new physics relevant to the discrepancy must couple to the muon and thus should leave traces in muon-antimuon collisions. 
The new particles in the reachable energy scales can be produced in the collider and thus the $g-2$ can be tested indirectly by searching for the new particles. If heavy particles are so heavy that they are not reachable, or if there exist just various higher dimensional operators without new particles, the process
 $\m\bar{\m}\to h \gamma$, where $h$ is a Higgs boson and $\gamma$ is a monochromatic photon, via a dimension six operator is a robust prediction. 
This process has a suppressed SM background.
 Consequently, the new physics contribution to the muon $g-2$ can be measured at the muon collider by measuring this cross section. 

A muon collider test of the $g-2$ was studied by the authors of Ref.~\cite{Capdevilla:2020qel}, by focusing on the lightest new particle, which is assumed to be reachable at the muon collider. 
In this Letter, we will show that even if none of new particles is reachable, the muon $g-2$ can be tested. 
In the reachable case, on the other hand, we provide complementary approaches, e.g. measuring the charged decay products of a heavier new particle. This should be the simplest way, in certain cases, e.g. if the lightest one does not have charge, and the heavier one is reachable. Also, from the properties of the produced new particles, we may ``measure" the $g-2$ by calculating the corresponding loop contribution.

\section{Muon $g-2$ from GeV-TeV physics and muon collider}

Let us consider a general renormalizable UV theory. 
The $g-2$  operator \eq{g-2} can be generated through loop diagrams including new states. At the 1-loop level\footnote{At higher loop generation, 
the production cross section at the muon collider should be even enhanced and the scenario is easier to be tested. The tree level process is not important in renormalizable UV theory. }, the diagram can be either composed of the following set of particles:
\beq\laq{cases} ( X_1, P_i^{\rm SM}) \OR (X_1, X_2)\eeq 
where 
$X_i \AND P_i^{\rm SM}$ denotes new particles and SM particles, respectively. 
The sum of the electromagnetic charges of the two particles is $-1$, which is the charge of a muon.

\subsection{A general approach in renormalizable theory}

Before discussing in detail, let us make a general discussion. 
To test the former case, one may generally measure the cross section of the SM process, $\mu\bar{\mu}\to P_i ^{\rm SM} \bar{P}^{\rm SM}_i$, which is from a $X_1$ mediated diagram.  (If $P_{i}^{\rm SM}$ is a neutrino, the discussion would be almost the same as the second case.)
In the context of $X_1=Z'$, 
it was shown in  Ref.~\cite{Baek:2001kca,Capdevilla:2020qel}  that the $Z'$ mediated process would lead to a significant enhancement of the scattering cross section of $\mu\bar\mu\rightarrow \mu\bar\mu$, 
and that the muon collider should be possible to discriminate the scenario from the SM.\footnote{This is the case that $Z'$ is heavy enough. In fact a light $Z'$ can also explain the muon $g-2$ anomaly with a small coupling, which is an IR scenario and is not our focus.  This case may not be tested at the muon collider (See, on the  other hand, the $Z'$ search in DUNE, M$^3$, and NA64~\cite{Kahn:2018cqs, Gninenko:2018tlp,Ballett:2019xoj}.)}

Another model-independent discussion, which applies to both cases, should be the search of the products of a heavier new particle decay. 
In the remainder of this subsection, we use the notation of the latter case in \eq{cases}, but our discussion holds for the former case by replacing $X_1$ to be $X$ and $X_2$ to be $ P_i^{\rm SM}.$
Suppose the typical coupling of the new particle is $g$ and the heaviest new particle in the loop is $X_1$, whose mass is $M_X.$
The muon $g-2 $ anomaly contribution can be denoted in the form
\beq
\d \a_\m=   \k \frac{m_\m^2 g^2}{16\pi^2 M_X^2}
\eeq
where $\k$ is a  model-dependent function of parameters. When $\k=\O(1)$, the UV contribution to the \eq{eff}  is 
proportional to the muon Yukawa coupling, $y_\mu$, which is satisfied in various models. 
$\k>\O(1)$ does not change our conclusion. 

To explain the discrepancy, 
\beq
M_X= 340\GEV \times g \sqrt{\k} \sqrt{ \frac{2.7\times 10^{-9}}{\d \a_\m}}.\laq{Mass}
\eeq
Therefore, even if $g \sqrt{\k}=\sqrt{4\pi}$ which is around the perturbative unitarity bound if $\k=\O(1)$, the 
mass of the new state is around $1\TEV,$ and thus is reachable in the muon collider of center-of-mass energy $\O(1-10)\TEV. $
The question is whether the heavy state can be significantly produced. 
In fact, by cutting the heavy state propagator in the loop for the muon $g-2$, one gets the $\mu +\bar{\mu} \to X_1 +\bar{X}_1$ process. 
The corresponding cross section can be estimated as 
\beq
 \simeq \k_2\frac{g^2 }{4\pi  E^2_{\rm cm}} ~({\rm for}~ E_{\rm cm}>2M_X).
\eeq
Here $\k_2$ is another model-dependent function of momentum and parameters, and $E_{\rm cm}$ is the center-of-mass energy.
By substituting \eq{Mass}, the lower band of the cross section can be obtained as  
\beq
\laq{cro}
\sigma \gtrsim  230 \,{\rm pb}\times \frac{\k_2}{\k^2} \(\frac{M_X}{310\GEV}\)^4 \(\frac{500\GEV}{E_{\rm cm}}\)^2   \(\frac{\d\a_\mu}{2.7\times 10^{-9}}\).
\eeq
Here we take the inequality because 
there could be other processes like a Drell-Yan production of $X_1,\bar{X}_1$ pair if $X_1$ is charged. 
By assuming a year of the run and $10^7$ s/year operation of a muon collider at the center-of-mass energy around $M_X$~\cite{Delahaye:2019omf}, the number of events for $X_1, \bar{X}_1$ pair production is
\beq
N \simeq  2.3\times 10^7 \(\frac{\sigma}{{27\,\rm pb}}\)\(\frac{L_{i}}{10^{34}{\rm cm}^2s^{-1} \times {\rm 1 ~yr}}\).
\eeq
where $L_{ i}$ is an integrated luminosity. 
If $X_1$ does not carry any charge, $X_1$ soon decays to a muon and a charged new particle. This process should be searched for analog to the dilepton search in the context of $Z'$~\cite{Aaboud:2017buh, Sirunyan:2018exx}.\footnote{Even if the $X_1$ decays to neutral particles dominantly in a dark sector, the scenario may be tested from the mono-photon search {c.f.} Ref.\cite{Han:2020uak}. One can also test this case by producing and detecting the lighter charged particle~\cite{Capdevilla:2020qel}. } 
If $X_1$ is charged, on the other hand, it may decay into a muon and a neutral state\footnote{We may also have more exotic decay products if $X_1$ dominantly couples to them.  In any case we have charged decay product, and our conclusion does not change. }
So the event should be $$\mu+\bar{\mu}\to X_1 +\bar{X}_1\to \mu +\bar{\mu}+ X_2+\bar{X}_2,$$
where $X_2$ is assumed to be lighter than $X_1$.  
 The dominant background, e.g. if $X_2$ is neutral, is $$\mu+\bar{\mu}\to W^+ + W^- \to \mu +\bar{\mu}+\nu_\mu+ \bar{\nu}_\mu.$$ 
Notice that in the muon collider, the center-of-mass colliding energy is given unlike the hadron collider. Thus, one may study the kinematics of the out-going muons to identify the masses of the new particles, e.g.
 for the background case, two out-going muons are almost back to back. 
This is in contrast with the new physics case: since $X_1$ is massive, the produced muons is not so back to back. 
In any cases, the acoplanar dileptons in excess of expectations of $WW$ and $ZZ$ production may be the signal a la slepton searches in lepton colliders~\cite{Abe:2001gc, Weiglein:2004hn}.
 (For comparison the SM cross section of $\mu \bar{\mu}\to \mu \bar{\mu} +{\rm missing}$ is below a few fb with the center-of-mass energy $\sim 10 \TEV$~\cite{Han:2020uak}.) 

It is noteworthy to mention that one may measure the masses and spins of $X_1,\AND X_2$, and the renormalizable couplings to muon via the production cross sections, and thus one may reconstruct the $g-2.$ This will be discussed elsewhere.

\subsection{Testing muon $g-2$ in a muon-smuon-bino like system}
In this part, we discuss the latter case of \eq{cases} in a concrete model. 
Let us consider
\beq
{\cal L_{\rm int}}= -g_1 \bar{\mu}_L \hat{P}_L \f_R \l-g_2 \bar{\mu}_R \hat{P}_R \f_L \l+ \d M^2  \f_L^* \f_R+h.c. 
\eeq
where $\l$ is a Majorana fermion with mass $M_\lambda$ while $\f_{L,R}$ are complex scalars with mass squares $m_{L,R}^2$; $\mu_L$ and $\mu_R$ represent the left and right-handed muon, respectively.
This Lagrangian with restricted parameter relations can be identified as the one from the minimal supersymmetric SM (MSSM), in which $\l$ and $\f_{ L,R}$ are identified as bino and smuons, respectively.  
In the context of MSSM, the so-called Higgs mediation mechanism~\cite{Yamaguchi:2016oqz}\footnote{This is named in Ref.\,\cite{Yin:2016shg}.}, which is proposed by the present authors, can lead to the sizable contribution to the muon $g-2$ with this effective Lagrangian. 
For scenarios involving the Higgs mediation in explaining the muon $g-2$ anomaly, see Refs.~\cite{Yamaguchi:2016oqz, Yin:2016shg, Yanagida:2018eho, Yanagida:2020jzy, Nagai:2020xbq}, and explaining both muon and electron $g-2$ anomaly~\cite{Endo:2019bcj, Badziak:2019gaf}.\footnote{Also see Ref.~\cite{Yanagida:2019evh} for the $E_7/\SU(5)\times \U(1)^3$ unification of family.  In the scenario the mass degeneracy between bino and wino is ``predicted" to be within $\O(1)\%$. Thanks to the Higgs mediation, the bino can be the dominant dark matter with a bino-wino coannihilation, and, moreover, the bottom-tau Yukawa coupling unification can be achieved up to $\O(1)\%.$}

The muon $g-2$ contribution is calculated as (c.f. Ref.~\cite{Moroi:1995yh})
\beq
\d a_\mu \simeq 
- {g_1 g_2 \over 8\pi^2}{ m_\mu \d M^2\, M_\l \over  m_{L}^2 m_{R}^2}\,
 f\left( \frac{m_{L}^2}{M_\l^2}, \frac{m_{R}^2}{M_\l^2} \right), 
 \laq{gm2}
\eeq 
where we have used the mass insertion approximation on the scalar mixings of $\d M^2$ and 
$
f(x,y)=xy \(\frac{-3+x+y+xy}{(x-1)^2(y-1)^2}+\frac{2x\log{x}}{(x-y)(x-1)^3}-\frac{2y \log y}{(x-y)(y-1)^3}\).
$
This satisfies $0<f(x,y)<1 \AND f(1,1)=1/6$.
$\d M^2$ should satisfy  $$\d M^2\lesssim \min{[m_L^2,m_R^2]} .$$
Otherwise not only our approximation of mass insertion in \eq{gm2} is invalid but also there are dangerous tachyonic scalars or vacuum instability (c.f. Ref~\cite{Endo:2013lva}).

Let us check the minimal production cross section of the new particles at the muon collider. 
By fixing $g_1, g_2$, $\d\a_\m$ becomes the largest when $M_\l\sim m_L\sim m_R \sim \sqrt{|\d M^2|}.$ 
With the relation we obtain the maximum contribution as
\beq
\laq{upb}
|\d a_\mu| \lesssim \frac{\ab{g_1 g_2}}{8\pi^2}\frac{m_\mu}{M_\l}f_N\simeq 2\times 10^{-9} \(\frac{|g_1 g_2|}{0.01}\)\(\frac{1\TEV}{ M_\l}\) \(\frac{f_N}{1/6}\).
\eeq
We then get
\beq
\k \lesssim \ab{\frac{M_\l}{m_{\rm \mu}}}.
\eeq
The upper limit corresponds to the smallest production cross section via the same coupling. 
In this scenario, the Drell-Yan production of charged leptons of $\f_L$, or $\f_R$, is significant. The cross-section purely mediated by a off-shell photon is given as 
\beq
\sigma^{\rm Drell\text{-}Yan}_{\mu\bar\mu\to \phi_{L,R} \phi^*_{L,R}}\simeq \frac{e^4 }{24\pi E_{\rm cm}^2} \sim 0.5 {\, \rm fb}\(\frac{10\TEV}{E_{\rm cm}}\)^2
\eeq
By taking into account the $Z$-boson contribution, the cross section can differ by $\O(1)$ factor. 
Since the ``sleptons" can be produced with $L_i=\O(1) \text{ab}^{-1}$ with $E_{\rm cm}\lesssim 100\TEV$, they can be easily detected due to the clean environment like the case of electron-positron collider (e.g. \cite{Abe:2001gc, Weiglein:2004hn, Endo:2013lva}). 

If $m_{L}\OR m_R\ll M_\l$ or $M_\l\ll m_{L}, m_R$, we can show that $m_L$ or $m_R$ tends to be lighter than the degenerate case, and thus it should be easier to test via Drell-Yan process.

In this section we have assumed that the new states relevant to the muon $g-2$ are reachable at the muon collider. 
If $|g_1 g_2|\gtrsim \O(1)$ in \Eq{upb}, the new physics scale may be above $100\TEV$.
Then the new particles are difficult to reach at the muon collider. 
Nevertheless,  one can test the muon $g-2$   by directly measuring the $g-2$ as shown soon.

\section{Measuring muon $g-2$ in effective theory }

Now let us assume the possibility that no new physics state is reachable at the muon collider. 
The $g-2$ operator is directly related with the coupling to the Higgs boson, $h$, as\footnote{In the symmetric phase, we may also have a vertex of muons, (charged) Higgs, and $Z$ ($W$)  boson. This is more or less model-dependent, and we do not consider. }
\beq
\laq{eff}
\Delta {\mathcal L}_{\rm eff}\supset { y_\mu \over M^2}\frac{h}{\sqrt{2}} \bar{\m}\s_{\m\n}   F^{\m \n}
{\m} ,
\eeq
where we have used $m_\m =y_\mu v$ with $v\simeq174\GEV$ being the expectation value of the Higgs field, $h,$ and have replaced $v$ to be $v+h /\sqrt{2}$ from \Eq{g-2}.\footnote{One may also consider dimension $>6$ terms to generate \eq{g-2} with several Higgs fields. In this case, still we get a similar dimension $6$ term with several Higgs field replaced by the vev, and the cross section of $\mu \bar{\mu} \to h \g$ can be enhanced by $\O(1-10)$.   However we may get a much enhanced production cross-section of multiple Higgs boson and a photon at the high energy.  That said, in a natural setup the contribution from the dimension six term should be the dominant. }
Here, to explain the anomaly \eq{g2m}, 
\beq
M\simeq  7.6\TEV \(\frac{2.7\times 10^{-9}}{\D \a_\m}\)^{1/2}.
\eeq
Notice that \eq{eff} is a dimension-six term which becomes stronger in the collisions of muon-antimuon  at higher energy.

\begin{figure}[!t]
\begin{center}  
   \includegraphics
[width=105mm]{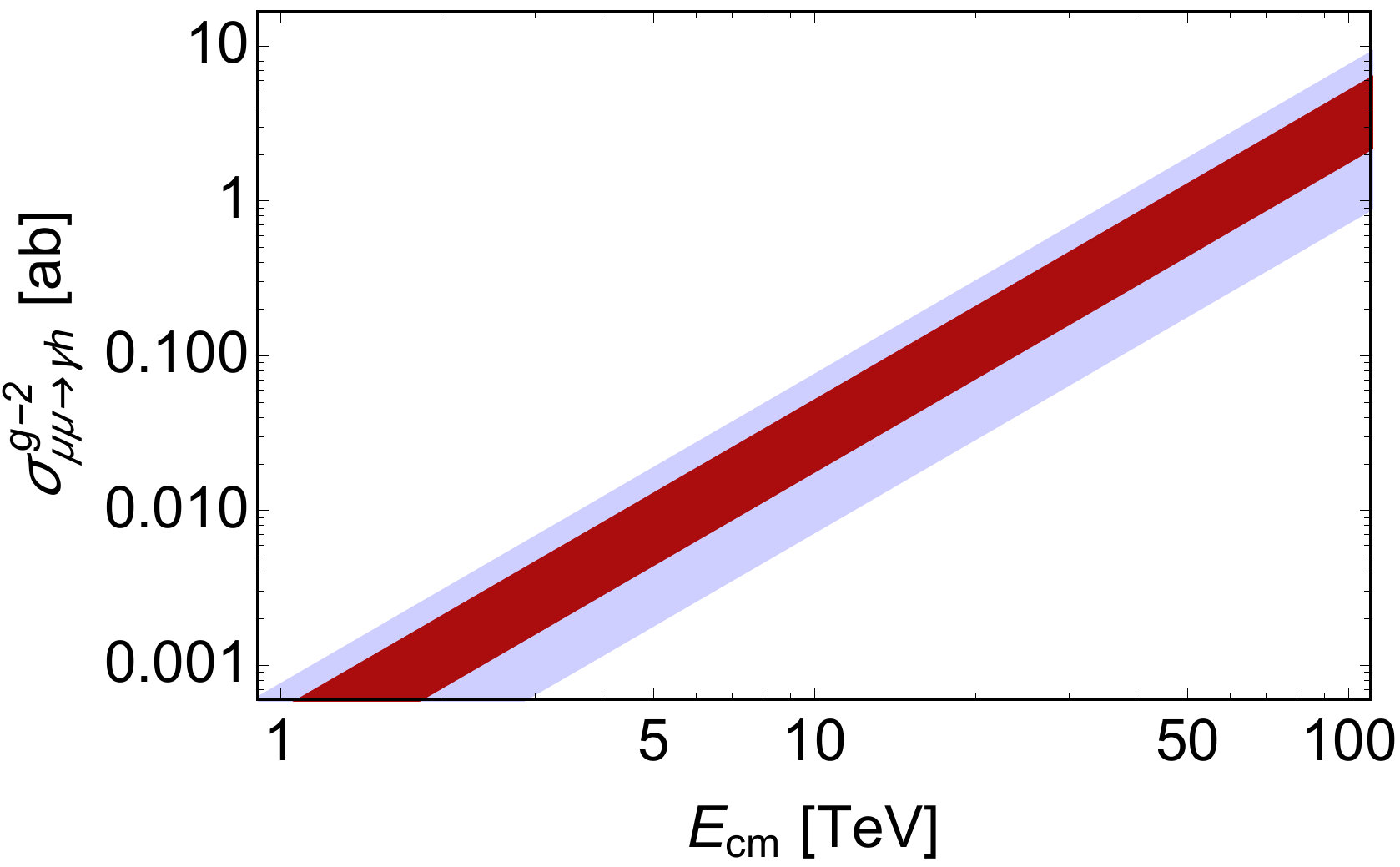}
      \end{center}
\caption{The cross section of $\m \bar{\m}\to h \g$ via the muon $g-2$ operator \eq{eff} by varying the center-of-mass energy. The red (blue) band corresponds to the $1\s$ ($2\s$) region of the muon $g-2$ anomaly. 
}\label{fig:1}
\end{figure}

The robust prediction of the muon $g-2$ anomaly in the effective theory is the enhancement on the $h$-$\g$ production of \beq \m +\bar{\m}  \to  h + \g\eeq 
at high energy.  
This provides a signal of a monochromatic photon with energy $\sim E_{\rm cm}/2$ and two fermions or four fermions with the invariant mass of the Higgs boson. 
The production cross section of $h \g$ can be calculated as 
\beq
\sigma^{g-2}_{\m\m\to \g h}\simeq \frac{y_\mu^2 E_{\rm cm}^2}{48\pi M^4}\sim 0.1 \text{ab}\(\frac{E_{\rm cm}}{20\TEV}\)^2\(\frac{7.6\TEV}{M}\)^4
\eeq
where we have neglected the masses of the initial and final states. 
The cross section at the tree-level is plotted in Fig.\,\ref{fig:1}. One finds that the cross section can be $\O(0.1-1)$ab $E_{\rm cm}= \O(10)\TEV$. 
The process $ \m +\bar{\m}  \to  h + \g$ is absent in the SM at the tree-level with neglecting the muon mass. 
At the loop level, for instance, the top-loop-induced contribution is
\beq \s^{\rm SM}_{\m\m\to \g h}\sim \frac{ y_t^2 e^6 }{(4\pi)^5 } \frac{1}{E_{\rm cm}^2}= 0.002\,{\rm ab} \(\frac{20\TEV}{E_{\rm cm}}\)^2.\eeq
Thus this can be safely neglected at a multi-TeV muon collider. 
The background of $\m\bar{\m}\to Z \g$ or $\m \bar{\m} \to W^+ W^- \gamma$ is suppressed with multi-TeV of $E_{\rm cm}$,\footnote{In principle, we can also test the $\m\bar{\m}\to h\g$ process in renormalizable UV models for the muon $g-2$. In this case, the cross-section is smaller at higher $E_{\rm cm}$ if the new particle masses are much smaller than $E_{\rm cm}. $
}  since the cross section decreases with larger $E_{\rm cm}$ (c.f. Refs.\,\cite{Kanemura:2018esc, Aoki:2019sht, Aoki:2020tba}). 
Consequently, one can test the muon $g-2$ by directly measuring the $g-2$ operator \eq{eff}.

\section{Conclusions}
We have shown that  the muon $g-2$ anomaly explanation by integrating heavy new states or with dimension six operator can be fully tested at a muon collider with center-of-mass energy up to $\O(10)\TEV$ and 
integrated luminosity of $\O(10){\rm ab}^{-1}$. A  model-independent approach was proposed, i.e. to produce the heaviest particle in the loop for the $g-2$ and measure its further decay into charged particles.
If the heavy particle masses are heavier than the reach of $\O(10)\TEV$ or there exist no heavy particles but just higher dimensional terms, we can, instead, measure $\m\bar{\m}\to h\g,$ whose cross section is significantly enhanced due to the nature of the higher dimensional operator. 

In our proposal the QCD non-perturbative effects are no more relevant, and our theoretical estimation is not bothered by them. 
Even if the muon $g-2$ anomaly was due to the wrong understanding of the QCD non-perturbative effects, the muon collider can give a direction to get a deeper understanding on the non-perturbative effects. 
More detailed study on collider phenomenology will be given in our future work. 
\\

{\it Note added:} While completing this paper, we found Ref. \cite{Buttazzo:2020eyl} appeared soon before our submission.
The authors studied the test of various higher dimensional terms, including the operator \eq{eff}, at the muon collider in the context of the muon $g-2$ anomaly.
Their discussions on \eq{eff} is consistent with ours.  Instead of focusing on other higher dimensional operators than \eq{eff}, we have proposed a generic approach to renormalizable models. 

\section*{Acknowledgments}
WY would like to thank the particle and cosmology group at Tohoku University for the kind hospitality when this project was initiated. 
WY also thanks M. Endo, J. Hisano, T. Moroi,  and M. Nojiri for useful discussions. 
This work was supported by JSPS KAKENHI Grant No.16H06490 (WY and MY) and 19H05810 (WY).

\providecommand{\href}[2]{#2}\begingroup\raggedright\endgroup

\end{document}